\documentstyle[12pt,epsfig]{article}

\catcode`\@=11
\long\def\@makefntext#1{
\protect\noindent \hbox to 3.2pt {\hskip-.9pt  
$^{{\ninerm\@thefnmark}}$\hfil}#1\hfill}		

\def\@makefnmark{\hbox to 0pt{$^{\@thefnmark}$\hss}}  
	
\def\ps@myheadings{\let\@mkboth\@gobbletwo
\def\@oddhead{\hbox{}
\rightmark\hfil\ninerm\thepage}   
\def\@oddfoot{}\def\@evenhead{\ninerm\thepage\hfil
\leftmark\hbox{}}\def\@evenfoot{}
\def\sectionmark##1{}\def\subsectionmark##1{}}

\renewcommand{\thefootnote}{\fnsymbol{footnote}}

\newcounter{sectionc}\newcounter{subsectionc}\newcounter{subsubsectionc}
\renewcommand{\section}[1] {\vspace*{0.6cm}\addtocounter{sectionc}{1} 
\setcounter{subsectionc}{0}\setcounter{subsubsectionc}{0}\noindent 
	{\normalsize\bf\thesectionc. #1}\par\vspace*{0.4cm}}
\renewcommand{\subsection}[1] {\vspace*{0.6cm}\addtocounter{subsectionc}{1} 
	\setcounter{subsubsectionc}{0}\noindent 
	{\normalsize\it\thesectionc.\thesubsectionc. #1}\par\vspace*{0.4cm}}
\renewcommand{\subsubsection}[1]
{\vspace*{0.6cm}\addtocounter{subsubsectionc}{1}
	\noindent {\normalsize\rm\thesectionc.\thesubsectionc.\thesubsubsectionc. 
	#1}\par\vspace*{0.4cm}}

\newcounter{appendixc}
\newcounter{subappendixc}[appendixc]
\newcounter{subsubappendixc}[subappendixc]

\renewcommand{\appendix}[1] {\vspace*{0.6cm}
        \refstepcounter{appendixc}
        \setcounter{figure}{0}
        \setcounter{table}{0}
        \setcounter{equation}{0}
        \renewcommand{\thefigure}{\Alph{appendixc}.\arabic{figure}}
        \renewcommand{\thetable}{\Alph{appendixc}.\arabic{table}}
        \renewcommand{\theappendixc}{\Alph{appendixc}}
        \renewcommand{\theequation}{\Alph{appendixc}.\arabic{equation}}
        \noindent{\bf Appendix \theappendixc #1}\par\vspace*{0.4cm}}

\def\abstracts#1{{
	\centering{\begin{minipage}{12.2truecm}\footnotesize\baselineskip=12pt\noindent
	\centerline{\footnotesize ABSTRACT}\vspace*{0.3cm}
	\parindent=0pt #1
	\end{minipage}}\par}} 

\renewenvironment{thebibliography}[1]
	{\begin{list}{\arabic{enumi}.}
	{\usecounter{enumi}\setlength{\parsep}{0pt}
\setlength{\leftmargin 1.25cm}{\rightmargin 0pt}
	 \setlength{\itemsep}{0pt} \settowidth
	{\labelwidth}{#1.}\sloppy}}{\end{list}}

\newcounter{itemlistc}
\newcounter{romanlistc}
\newcounter{alphlistc}
\newcounter{arabiclistc}

\newcommand{\fcaption}[1]{
        \refstepcounter{figure}
        \setbox\@tempboxa = \hbox{\footnotesize Fig.~\thefigure. #1}
        \ifdim \wd\@tempboxa > 6in
           {\begin{center}
        \parbox{6in}{\footnotesize\baselineskip=12pt Fig.~\thefigure. #1}
            \end{center}}
        \else
             {\begin{center}
             {\footnotesize Fig.~\thefigure. #1}
              \end{center}}
        \fi}

\newcommand{\tcaption}[1]{
        \refstepcounter{table}
        \setbox\@tempboxa = \hbox{\footnotesize Table~\thetable. #1}
        \ifdim \wd\@tempboxa > 6in
           {\begin{center}
        \parbox{6in}{\footnotesize\baselineskip=12pt Table~\thetable. #1}
            \end{center}}
        \else
             {\begin{center}
             {\footnotesize Table~\thetable. #1}
              \end{center}}
        \fi}

\def\@citex[#1]#2{\if@filesw\immediate\write\@auxout
	{\string\citation{#2}}\fi
\def\@citea{}\@cite{\@for\@citeb:=#2\do
	{\@citea\def\@citea{,}\@ifundefined
	{b@\@citeb}{{\bf ?}\@warning
	{Citation `\@citeb' on page \thepage \space undefined}}
	{\csname b@\@citeb\endcsname}}}{#1}}

\newif\if@cghi
\def\cite{\@cghitrue\@ifnextchar [{\@tempswatrue
	\@citex}{\@tempswafalse\@citex[]}}
\def\citelow{\@cghifalse\@ifnextchar [{\@tempswatrue
	\@citex}{\@tempswafalse\@citex[]}}
\def\@cite#1#2{{$\null^{#1}$\if@tempswa\typeout
	{IJCGA warning: optional citation argument 
	ignored: `#2'} \fi}}

 1
 1
 1

\font\ninerm=cmr9

\newcommand{\np}[3]{Nucl.\ Phys.\ {\bf B#1} (19#2) #3}

\def\simgt{\rlap{\lower 3.5 pt \hbox{$\mathchar \sim$}} \raise 1pt \hbox {$>$}}
\def\simlt{\rlap{\lower 3.5 pt \hbox{$\mathchar \sim$}} \raise 1pt \hbox {$<$}}

\newcommand{\beq}{\begin{equation}}
\newcommand{\eeq}{\end{equation}}
\newcommand{\bea}{\begin{eqnarray}}
\newcommand{\eea}{\end{eqnarray}}

\newcommand{\tb}{\mbox{tg$\beta$}}
\newcommand{\lsim}{\raisebox{-0.13cm}{~\shortstack{$<$ \\[-0.07cm] $\sim$}}~}

\def\msb{\overline{\rm MS}}

\def\squ{\tilde{Q}}
\def\tgb{\mbox{tg$\beta$}}
\def\hH{{\cal H}}
\def\beqn{\begin{eqnarray}}
\def\eeqn{\end{eqnarray}}

\catcode`@=11
\def\citer{\@ifnextchar [{\@tempswatrue\@citexr}{\@tempswafalse\@citexr[]}}

\def\@citexr[#1]#2{\if@filesw\immediate\write\@auxout{\string\citation{#2}}\fi
  \def\@citea{}\@cite{\@for\@citeb:=#2\do
       {\@citea\def\@citea{-}\@ifundefined
       {b@\@citeb}{{\bf ?}\@warning
       {Citation `\@citeb' on page \thepage \space undefined}}
       {\csname b@\@citeb\endcsname}}}{#1}}
\catcode`@=12

\begin{document}

\centerline{\normalsize\bf HIGGS BOSON PRODUCTION AT THE LHC}
\baselineskip=22pt

\centerline{\footnotesize MICHAEL SPIRA}
\baselineskip=13pt
\centerline{\footnotesize\it CERN, Theory Division, CH-1211 Geneva, Switzerland}
\baselineskip=12pt
\centerline{\footnotesize E-mail: spira@cern.ch}

\vspace*{0.9cm}
\abstracts{
Recent theoretical progress in the evaluation of Higgs boson production at
the LHC within the Standard Model and its minimal supersymmetric extension
is reviewed. In particular the two-loop QCD corrections to the squark loop
contributions to scalar Higgs production in the MSSM and soft gluon resummation
effects in Standard and SUSY Higgs production via the gluon fusion mechanism
are discussed.
}
 
\normalsize\baselineskip=15pt
\setcounter{footnote}{0}
\renewcommand{\thefootnote}{\alph{footnote}}

\section{Introduction}

The search for Higgs particles \cite{S:higgs} is one of the most
important endeavours for future high energy $e^+e^-$ and hadron
collider experiments.  The Higgs boson is the only particle of the
Standard Model (SM) which has not been discovered so far. The direct
search at the LEP1 experiments via the process $e^+e^- \to Z^* H$
yields a lower bound on the Higgs mass of $\sim 65.2$ GeV \cite{S:lep1}.
Theoretical consistency restricts the Higgs mass to be less than
$\sim 700$ GeV \cite{S:lattice}.  The dominant Higgs production
mechanism at the LHC, a $pp$ collider with a c.m.~energy of 14 TeV, is
the gluon fusion process $gg \to H$, which is mediated by a heavy quark
triangle loop at lowest order \cite{S:glufus}.  As an important step
to increase the theoretical precision, the two-loop QCD corrections
have been calculated, resulting in a significant increase of the
predicted total cross section by about 50--100\%
\cite{SDGZ,S:limit}. The dependence on the unphysical renormalization
and factorization scales decreased considerably by including these
next-to-leading-order (NLO) corrections, resulting in an estimate of
about 15\% for the remaining scale sensitivity \cite{SDGZ}.  It is
important to note that the NLO corrections
are dominated by soft gluon radiation effects.

The minimal supersymmetric extension (MSSM) is
among the most attractive extensions of the SM.  It requires the
introduction of two Higgs doublets leading to the existence of five
scalar Higgs particles: two scalar CP-even $h,H$, one pseudoscalar
CP-odd $A$ and two charged bosons $H^\pm$. This Higgs sector can be
described by two parameters, which are usually chosen to be
\tb, the ratio of the two vacuum expectation values, and the
pseudoscalar Higgs mass $M_A$. Including higher-order corrections to
the Higgs masses and couplings up to the two-loop level, the mass of
the lightest scalar Higgs particle $h$ is restricted to be smaller
than $\sim 130$~GeV \cite{S:hbound}. The direct search at LEP1 sets
lower bounds of about 45 GeV on the masses of the MSSM Higgs bosons
\cite{S:lep1}.  The dominant neutral Higgs production mechanisms at
the LHC are the gluon fusion processes $gg \to h,H,A$ and Higgs-strahlung off
bottom quarks, $gg,q\bar q \to b\bar b
h,b\bar b H,b\bar b A$, which become important only for large
\tb~\cite{S:phibb}. The coupling of the neutral Higgs particles to
gluons is again mediated by top and bottom loops, with the latter
providing the dominant contribution for large \tb, and squark loops,
if their masses are smaller than about 400 GeV \cite{S:squark}.
The two-loop (NLO) QCD
corrections to the quark loop contributions to the gluon fusion mechanism
have also been calculated
\cite{SDGZ,S:squark}, and conclusions completely analogous to the SM
case emerge.  Soft gluon radiation effects again provide the dominant
contribution to these corrections, for small \tb.

\section{Gluon Fusion: Squark Loops}

Recently the QCD corrections to the squark loop contributions to the
cross sections $\sigma(pp \to \hH + X)$ of the fusion processes
for the neutral CP-even Higgs particles $\hH = h,H$ 
\begin{equation}
gg \to \hH (g)~~~~\mbox{and}~~~~gq \to \hH q,~~ q\bar q \to \hH g
\label{eq:proc}
\end{equation}
have been calculated.
Because of CP invariance, squark loops do not contribute to the
production of the CP-odd Higgs boson in lowest order. The QCD
corrections from squark loops were evaluated in the heavy squark limit,
where the calculation can be simplified by extending the lowest-order
low-energy theorems \cite{SDGZ,S:limit,lowen} to two loops. This limit should
be a good approximation \cite{SDGZ} for the production of Higgs
particles with masses smaller than twice the squark masses.
For simplicity, we will restrict
ourselves to the case of degenerate squarks where mixing effects are
absent. Also in this case, scalar
squarks do not contribute to the production of the CP-odd
Higgs boson A at NLO.

The LO cross sections for CP-even Higgs production at the LHC are given
by 
\beq
\sigma_{LO}(pp \to \hH + X) = \sigma_0^{\hH} \tau_{\hH} \frac{d{\cal L}^{gg}}
{d\tau_{\hH}}
\eeq
with $d{\cal L}^{gg}/d\tau_{\hH}$ the gluon luminosity at
$\tau_{\hH}=M_{\hH}^2/s$ and $s$ the total c.m.~energy. The parton
cross sections are built up from  heavy quark and squark amplitudes, 
\beq
\sigma_0^{\hH} = \frac{G_F\alpha_s^2}{128\sqrt{2}\pi} \left| \sum_Q g_Q^{\hH}
\tau_Q \left[ 1+(1-\tau_Q) f(\tau_Q) \right]
- \sum_{\squ} g_{\squ}^{\hH} \frac{1}{2}\tau_{\squ} \left[ 1-\tau_{\squ}
f(\tau_{\squ}) \right]
\right|^2
\label{eq:sig0}
\eeq
with the scaling variables $\tau_{Q/ \squ}\equiv 4m_{Q/ \squ}^2/M_{\hH}^2$
and using the scalar triangle integral
\beq
f(\tau) = \left\{ \begin{array}{ll}
\displaystyle \arcsin^2 \biggl(\frac{1}{\sqrt{\tau}}
\biggr) & \tau \ge 1 \\
\displaystyle - \frac{1}{4} \left[ \log \frac{1+\sqrt{1-\tau}}
{1-\sqrt{1-\tau}} - i\pi \right]^2 & \tau < 1
\end{array} \right. \ .
\eeq
The normalized scalar  quark and squark couplings to the CP-even Higgs
bosons, $g^{\hH}_{Q,{\squ}}$, can be found in Ref.\cite{SDGZ}.
The sums run over $t,b$ quarks and the left- and right-handed
squarks $\squ_L, \squ_R$, which in the absence of mixing are identical
to the mass eigenstates.

The QCD corrections to the gluon fusion process, eq.~(\ref{eq:proc}),
consist of virtual two-loop corrections and one-loop real corrections
due to gluon radiation, as well as contributions from quark--gluon
initial states and quark--antiquark annihilation.  The renormalization
program has been carried out in the $\overline{\rm MS}$ scheme for the
strong coupling constant and the parton densities, while the quark and
squark masses are defined at the poles of their respective propagators.
The result for the cross sections can be cast into the form
\beq
\sigma(pp\to \hH +X) = \sigma^{\hH}_0 \left[1+C_{\hH}(\tau_Q, \tau_{\squ})
\frac{\alpha_s}{\pi}
\right] \tau_{\hH} \frac{d{\cal L}^{gg}}{d\tau_{\hH}}
+ \Delta \sigma^{\hH}_{gg}
+ \Delta \sigma^{\hH}_{gq} + \Delta \sigma^{\hH}_{q\bar q} \, .
\label{eq:sigstruc}
\eeq
The coefficient $C_{\hH}$ denotes the virtual two-loop corrections
regularized by the infrared singularities of the real gluon emission.
The terms $\Delta \sigma^{\hH}_{ij}$ ($i,j=g,q$) denote the finite parts
of the real corrections due to gluon radiation and the $gq$ and $q\bar
q$ initial states. The expressions for the $t,b$ quark contribution can
be found in Refs.\cite{SDGZ,S:limit}. 

The calculation of the QCD corrections has been performed by extending
the low-energy theorems \cite{SDGZ,S:limit,lowen} to scalar squarks at the
two-loop level.  For a light CP-even Higgs boson, these theorems
relate the matrix elements of the quark and squark contributions to the
Higgs--gluon vertex to the gluon two-point function. In the following we
consider
only the pure gluon exchange contributions, which are expected to be the
dominant ones; for heavy enough gluinos, the two-loop corrections due
to gluino exchange should be small, since they are suppressed by inverse
powers of the gluino mass.
The final result for the squark contributions to the $\hH$
coupling to gluons can be expressed in terms of the effective Lagrangian
[$v=(\sqrt{2}G_F)^{-1/2}$]
\beq
{\cal L}_{eff}^{\squ} = \sum_{\squ} \frac{g_{\squ}^{\hH}}{4}
\frac{\beta_{\squ}(\alpha_s)/\alpha_s}{1+\gamma_{\squ}(\alpha_s)} 
G^{a\mu\nu} G^a_{\mu\nu} \frac{\hH}{v} 
= \sum_{\squ} g_{\squ}^{\hH} \frac{\alpha_s}{12\pi}
G^{a\mu\nu} G^a_{\mu\nu}\frac{\hH}{v} \left[ 1+\frac{25}{6}\frac{\alpha_s}{\pi}
\right] \, ,
\label{eq:leff}
\eeq
where we have used the anomalous squark mass dimension
\cite{susybet}, $\gamma_{\squ} = 4\alpha_s/(3\pi)$,
and the squark contribution to the QCD $\beta$ function \cite{betafun},
$\beta_{\squ}(\alpha_s) = \alpha^2_s/(12\pi) \left[
1+ 11\alpha_s/(2\pi) \right]$.
Starting from the Lagrangian of eq.~(\ref{eq:leff}), the effective QCD
corrections
due to real gluon emission and the $gq/q\bar{q}$ initial states have to be
added. These corrections are identical to the corresponding corrections
to quark loops \cite{SDGZ} in the heavy quark limit. 

The QCD-corrected squark loop amplitudes have to be added coherently to
the corrected $t,b$ loop amplitudes, whose full mass dependence
is known. To obtain a more reliable prediction for the total
cross sections, the resulting amplitudes for the squark contributions
have been normalized to the lowest-order amplitude in the limit of large
squark masses. These ratios are then multiplied by the lowest-order
amplitude including the full squark mass dependence. The heavy squark
limit is then expected to be a very good approximation for Higgs masses
below the ${\squ} {\squ}^*$ threshold, as in the corresponding case of
top quark contributions \cite{SDGZ}. 
The final results for the partonic cross sections defined in
eq.~(\ref{eq:sigstruc}) can be found in \cite{S:squark}.

\begin{figure}[htb]
\vspace*{-3.0cm}
\hspace*{2.0cm}
\epsfig{file=kfactor.fg,bbllx=0pt,bblly=70pt,bburx=575pt,bbury=800pt,%
        width=10.0cm,angle=0}
\vspace*{-2.3cm}
\fcaption{\it
Ratio of the QCD-corrected cross section to the lowest order result for
$\sigma(pp\rightarrow \hH + X)$
for $\tgb = 1.5$ and 30. Solid lines include $t,b$ and  squark
contributions (with $m_{\squ}=200$~GeV); dashed lines include 
only the $t,b$ contributions.
We have taken
$m_b=5$ GeV, $m_t=176$ GeV and 
$\alpha_s(M_Z^2) = 0.118$. 
We use  next-to-leading
order GRV parton densities \cite{GRV}
and take the renormalization scale $\mu$
and the factorization scale $M$ equal
to $M_{h/H}$.}
\end{figure}

In Fig.~1, we present the $K$-factors for  the QCD corrections to the
production of the CP-even MSSM Higgs bosons as functions of the ${\hH}$
masses for the LHC at a c.m.~energy $\sqrt{s} = 14$ TeV with (solid
lines) and without (dashed lines) the squark contributions.
The $K$-factors are defined as the ratios of the QCD-corrected and lowest-order
cross sections, using next-to-leading order $\alpha_s$ and parton
densities in both terms. A common value $m_{\squ}= 200$ GeV has been
used for the left- and right-handed squark masses.
This value is identified with the SUSY scale of the MSSM couplings 
and Higgs masses.
The QCD corrections to the squark loops are large,
approximately of the same
size as the QCD corrections to the quark loops.
They enhance the cross sections by a factor between 1.6
and 2.8.  However, if the
lowest-order cross sections are convoluted with lowest-order $\alpha_s$
and parton densities, the $K$-factors are reduced to a level between 1 and 2.
It can be inferred from Fig.~1 that the inclusion of squark loops in the
production of both CP-even Higgs particles $h$ and $H$ does not
substantially modify the $K$-factors compared to the case where squark
loops are absent.
Thus, to a good approximation, the effect of the squark loops in the
gluon fusion mechanism is quantitatively determined by the lowest-order
cross section (including squark loop contributions), multiplied by the
known $K$-factors when only the $t,b$ quark contributions
\cite{SDGZ} are taken into account. 

\section{Soft Gluon Resummation}

As a next step a resummation of soft gluon radiation effects in
Higgs boson production via gluon fusion has been performed.
If the $K$-factor, evaluated in the heavy top quark limit, is multiplied by the
full massive lowest-order cross section, the result approximates the SM
gluon-fusion cross section at NLO within 10\% and the MSSM cross sections
within 25\% for $\tb\lsim 5$ \cite{hresum}. In the heavy quark limit the
partonic cross sections factorize as
\beq
\hat\sigma^\phi_{gg} = \sigma^\phi_0~\kappa_\phi~\rho_\phi
(z,M_\phi^2/\mu^2,\epsilon)
\label{rhodef}
\eeq
with the coefficient $\sigma_0^{h/H}$ defined in eq.~(\ref{eq:sig0}) and
\beq
\sigma^A_0 = \frac{G_F \alpha_s^2}{128\sqrt{2}\pi} \left| \sum_Q g_Q^A \tau_Q
f(\tau_Q) \right|^2 \, ,
\label{sigzero}
\eeq
where $g_Q^\phi (\phi = h,H,A)$ denote the
modified top Yukawa couplings normalized to the SM coupling, which are
given in \cite{SDGZ}. In the following we will neglect the $gq, q\bar q$
initial states, which contribute up to $\sim 10\%$ at NLO, and the squark loops,
by choosing the common squark mass as $m_{\squ}=1$ TeV.
The factor $\kappa_\phi$ in eq.~(\ref{rhodef})
stems from the effective coupling of the Higgs boson to gluons in the
heavy top quark limit,
which can be obtained by means of low-energy
theorems \cite{SDGZ,lowen}. They are
given\footnote{The factors $\kappa_{h,H}$ include the
  top quark contribution at {\it vanishing} momentum transfer, which
  differs from the top quark contribution to the $\overline{MS}$
  $\beta$ function by a finite amount at ${\cal O}(\alpha_s^4)$
  \cite{bernwetz}.}~
by \cite{hresum}
\bea
\kappa_{h,H} & = & 1 + \frac{11}{2} \frac{\alpha_s(m_t^2)}{\pi}
+ \frac{3830 - 201\, n_f}{144}\left( \frac{\alpha_s(m_t^2)}{\pi}\right)^2
\nonumber \\
& & + \frac{153-19\, n_f}{33-2\, n_f}
~\frac{\alpha_s(M_{h,H}^2) - \alpha_s(m_t^2)}{\pi}
+ {\cal O}(\alpha_s^3) \\
\kappa_A & = & 1 \, ,
\eea
where $\alpha_s$ is the strong coupling constant in the $\msb$ scheme
including $n_f=5$ flavours in the evolution, i.e.\ the top quark is decoupled.
The scale of the higher-order corrections to this effective coupling has to
be identified with the top quark mass $m_t$ \cite{SDGZ,lowen}.

We will now construct a resummed expression for the correction factors
$\rho_\phi$ by means of the methods
described in Ref.~\cite{CLS}.  Near the elastic edge of phase space
the Higgs cross section in the infinite mass limit may be factorized
into hard, soft and jet functions, in complete analogy with the
Drell--Yan cross section. Following the arguments of Ref.~\cite{CLS}
this leads to the Sudakov evolution equation:
\beq
M_\phi^2 \frac{d}{dM_\phi^2} \rho_\phi \left(z,\frac{M_\phi^2}{\mu^2},
\alpha_s(\mu^2) \right)\! = \!
   \int^1_z \frac{dz'}{ z'} W_\phi \left(z',\frac{M_\phi^2}{\mu^2},
\alpha_s(\mu^2) \right)
\rho_\phi \left(\frac{z}{z'},\frac{M_\phi^2}{\mu^2},\alpha_s(\mu^2) \right)
\, .
\label{sudevz}
\eeq
The solution to the moments\footnote{We define the Mellin--moments in the
usual way as $\tilde f(N) = \int_0^1 dz z^{N-1} f(z)$.}~ of eq.~(\ref{sudevz})
is given by
\beq
\tilde \rho_\phi \left(N,\frac{M_\phi^2}{\mu^2},\alpha_s(\mu^2) \right)
= \exp\left[ \int_0^{M_\phi^2}
 {d \xi^2 \over \xi^2} 
{\tilde W}_\phi \biggl(N,{\xi^2\over \mu^2},\alpha_s(\mu^2)\biggr) 
\right] \, ,
\label{sudsolveN}
\eeq
which may be expanded perturbatively. In this way the one-loop coefficients
$W_\phi^{(1)}$ of the evolution kernel $W_\phi$ can be
straightforwardly determined from the explicit NLO calculation \cite{hresum}.
In the following we will define three approximations:
scheme $\alpha$ includes only the leading $\log N$ contributions, scheme
$\beta$ all terms of ${\cal O}(N^0)$, while scheme $\gamma$ contains terms of
${\cal O}(\log N/N)$ in addition, which arise from collinear gluon--gluon
splitting and are thus universal. [However, they are not covered by the
present level of factorization theorems \cite{factheo}, which leads to
eq.~(\ref{sudevz}).]

In order to obtain a finite expression for the correction factors $\rho_\phi$
we now have to renormalize the strong coupling constant and perform mass
factorization of the gluon distributions. Both objects have been defined in the
$\overline{MS}$ scheme.
The final results for the moments of the renormalized correction factors can
be found in \cite{hresum}. They have been expanded perturbatively up to NLO and
NNLO, thus yielding a quantitative estimate of the unknown NNLO corrections
to the gluon-fusion mechanism \cite{hresum}.

\begin{figure}[htb]

\vspace*{-0.8cm}
\hspace*{-0.8cm}
\epsfig{file=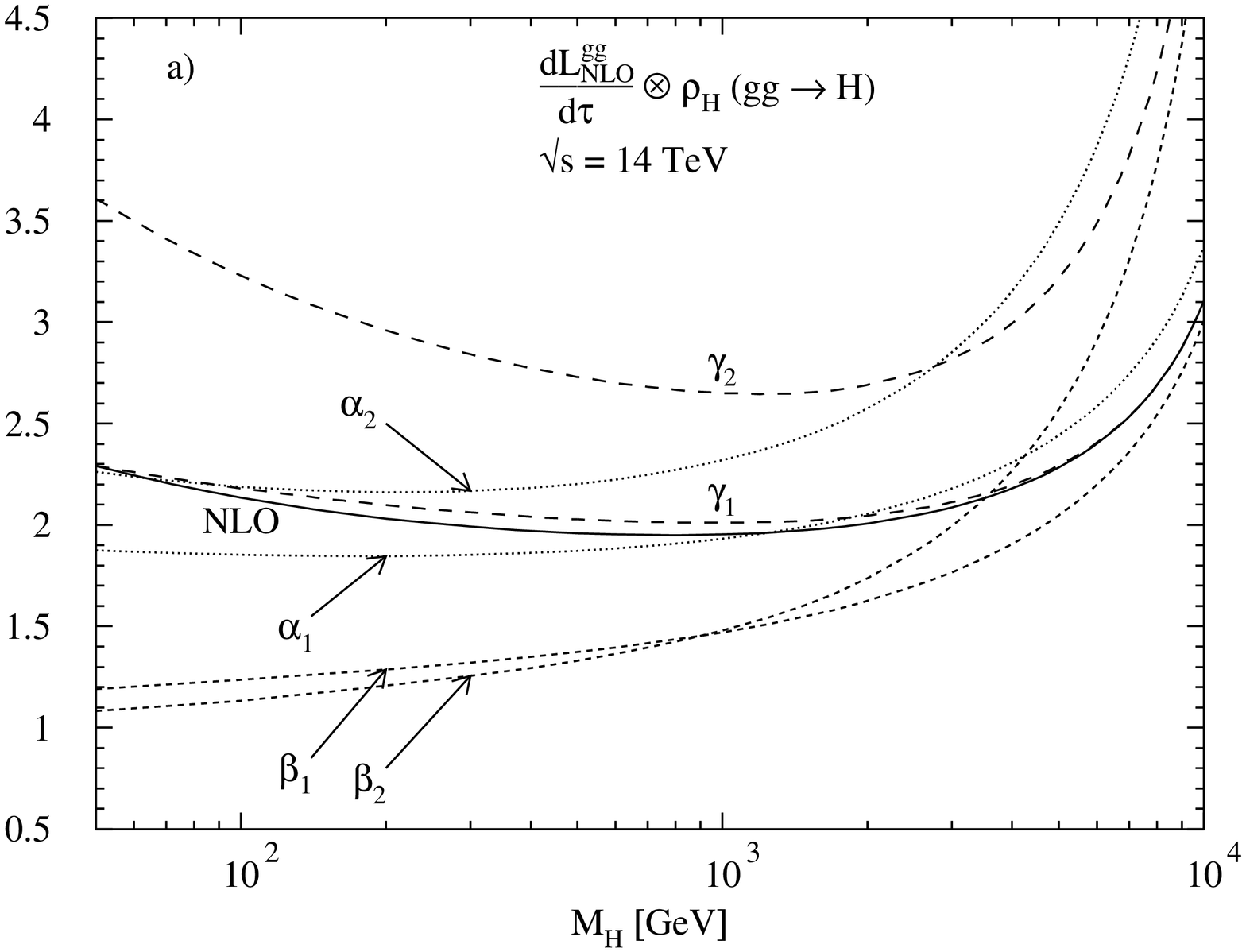,bbllx=0pt,bblly=70pt,bburx=575pt,bbury=800pt,%
        width=8.0cm,height=8.5cm,angle=-90}

\vspace*{-8.04cm}
\hspace*{7.0cm}
\epsfig{file=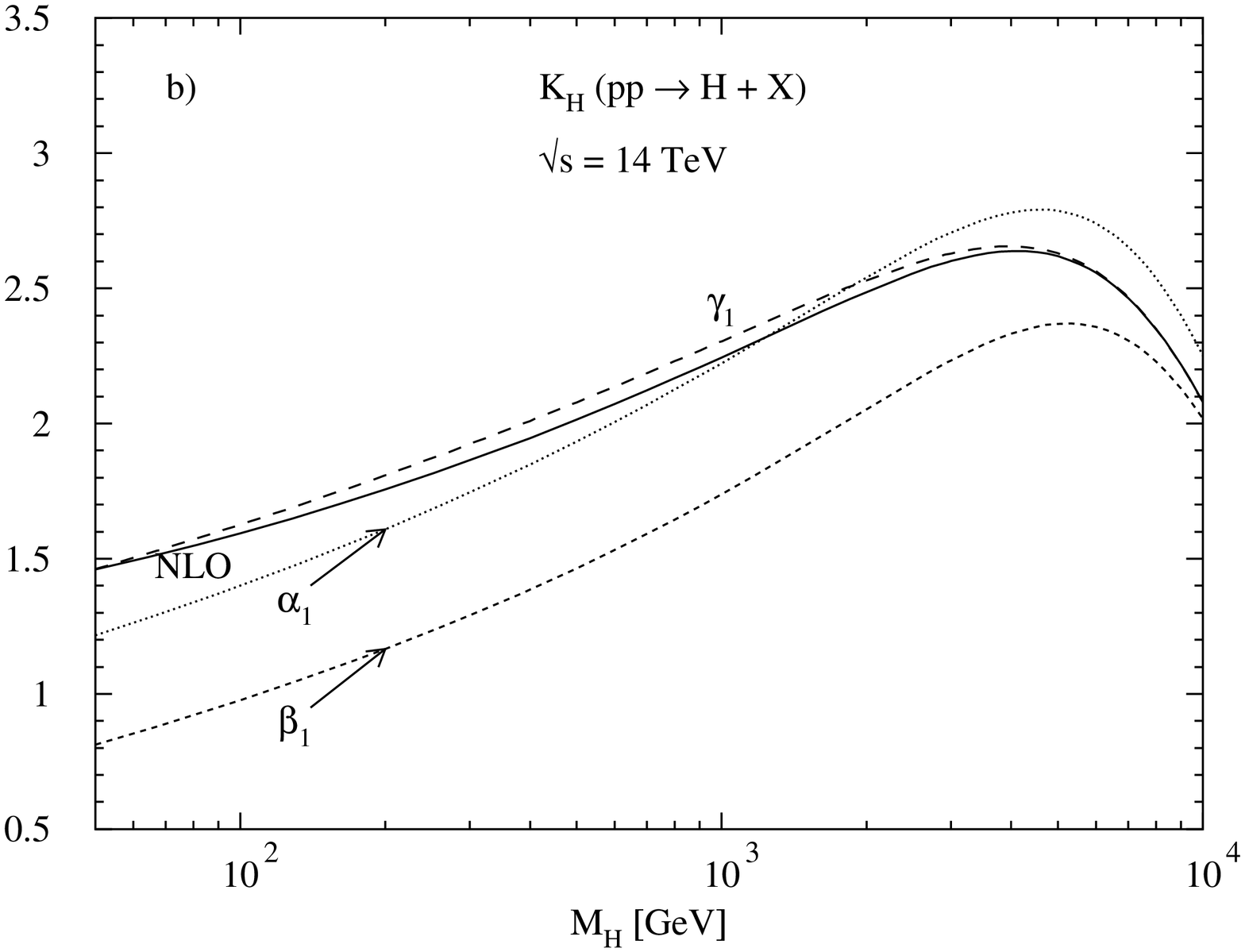,bbllx=0pt,bblly=70pt,bburx=575pt,bbury=800pt,%
        width=8.0cm,height=8.5cm,angle=-90}
\vspace*{-0.5cm}

\fcaption{\it a) Exact and approximate two- and three-loop correction
  factors, convoluted with the NLO gluon--gluon luminosity $d{\cal
    L}^{gg}_{NLO}/d\tau$, in the heavy top-mass limit. The results for
  the three different schemes are presented as a function of the
  scalar Higgs mass $M_H$, using NLO CTEQ4M parton densities
  \cite{CTEQ4} and $\alpha_s$ [$\Lambda_{\msb}^{(5)} = 202$ MeV].  b)
  Hadronic NLO K-factor using LO CTEQ4L parton densities \cite{CTEQ4}
  and $\alpha_s$ [$\Lambda_{\rm LO}^{(5)} = 181$ MeV] for the LO cross
  section and including the NLO contributions from $\kappa_H$.}
\label{Kf1}
\end{figure}
Fig.~\ref{Kf1}a shows the NLO and NNLO expansions of the scalar correction
factor, convoluted with NLO gluon densities, as a function of the Higgs mass.
A similar picture emerges for the pseudoscalar Higgs boson. The NLO
correction factor in scheme $\gamma$ coincides with the exact NLO result
within less than 5\%, while schemes $\alpha$ and $\beta$ fail to approximate
the exact NLO calculation reliably. From an analogous analysis of the
Drell--Yan process we get strong confidence that scheme $\gamma$ also
approximates the NNLO expansion with reliable accuracy \cite{hresum}. Thus
it might be expected that the curve labelled $\gamma_2$ yields a reasonable
approximation of the unknown NNLO corrections to the gluon-fusion process.
The NNLO correction factor amounts to 2.7--3.5 in the relevant Higgs mass
range, which is very large. However, the physical $K$-factor requires the LO
result to be convoluted with LO parton densities and strong coupling and the
NNLO expansion with NNLO quantities. This leads to a strong reduction of
the NLO $K$-factor in comparison to the naive correction factor, as can be
inferred from Fig.~\ref{Kf1}b. Due to the lack of NNLO parton densities no
physical prediction of the Higgs production cross section at NNLO can be
made. From the situation at NLO one might expect a significant decrease of
the NNLO $K$-factor compared to the result of Fig.~\ref{Kf1}a.

An alternative way to reach a consistent prediction of the gluon-fusion cross
section at NNLO may be provided by extracting the gluon luminosity from
another production cross section at the LHC, e.g.~$gg\to t\bar t$, the resummed
expression of which also has to be expanded up to NNLO. This result may then be
inserted in the convolution integral for the Higgs production cross section.
This method would thus allow us to obtain approximate NNLO parton densities for
physical predictions of production cross sections at the LHC. In the same way
we could also perform predictions of the full resummed results. This, however,
requires a prescription, how to regularize the infrared renormalon singularity
\cite{renorm}, which appears in the integrals of the running strong coupling
constant.

\begin{figure}[htb]

\vspace*{-0.5cm}
\hspace*{1.5cm}
\epsfig{file=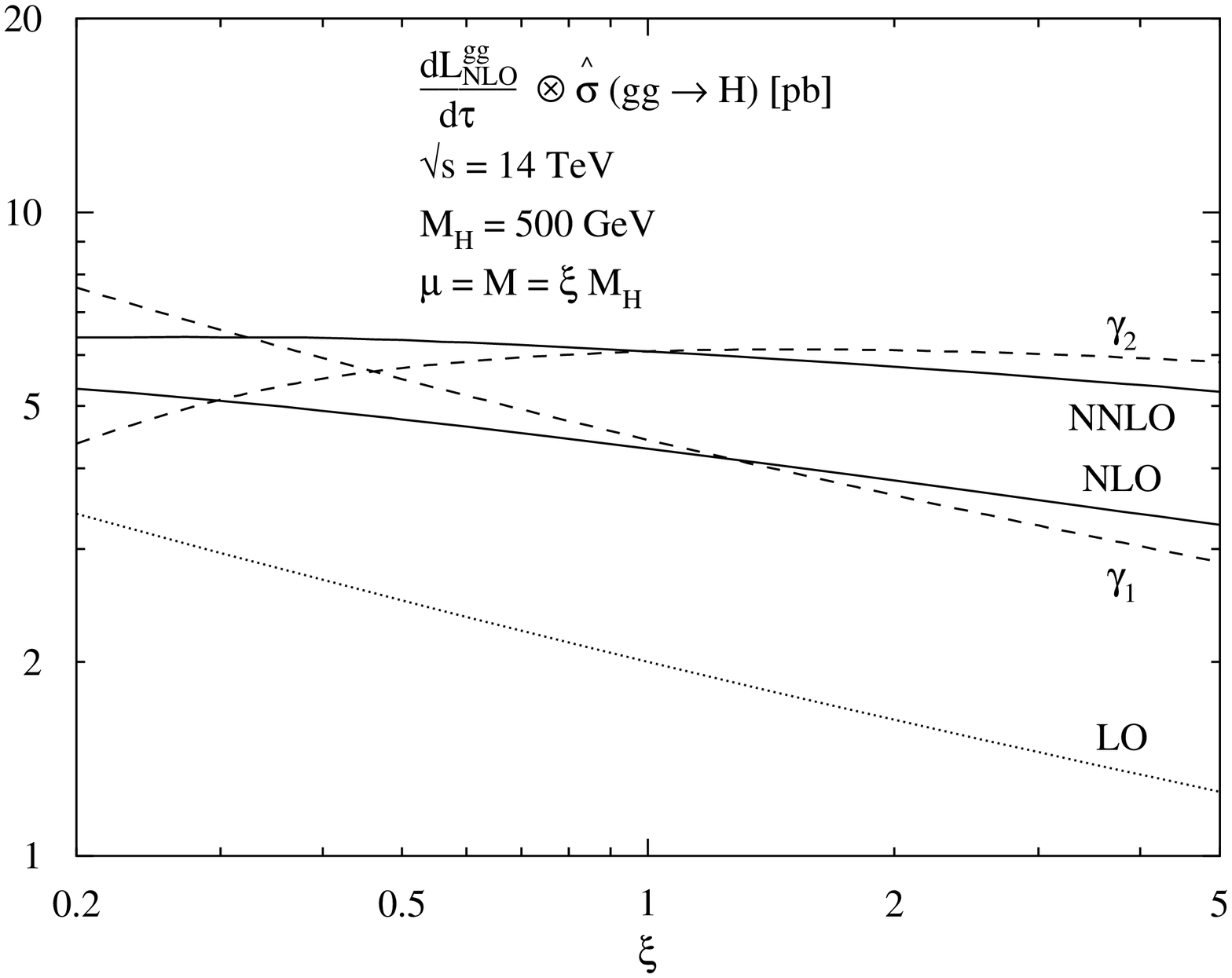,bbllx=0pt,bblly=70pt,bburx=575pt,bbury=800pt,%
        width=9cm,angle=-90}
\vspace*{-0.3cm}

\fcaption{\it Scale dependence of the Higgs production cross section for two
  values of the Higgs mass $M_H$.  NLO CTEQ4M parton densities
  \cite{CTEQ4} and strong coupling [$\Lambda_{\msb}^{(5)} = 202$ MeV]
  have been used in all expressions, so that the NNLO results do not
  correspond to the physical NNLO cross sections.}
\label{fg:Higgs_sc}
\end{figure}
Fig.~\ref{fg:Higgs_sc} presents the scalar Higgs production cross section at LO,
NLO and NNLO as a function of the renormalization/factorization scale in units
of the Higgs mass for $M_H=500$ GeV. Again, all orders of the cross section
have been evaluated with NLO parton densities and strong coupling, and thus the
NNLO curves do not correspond to the physical cross section at NNLO. The dashed
lines present the NLO and NNLO results of scheme $\gamma$, where the NNLO
contribution has been added to the exact NLO result. The solid NLO curve
corresponds to the exact NLO cross section in the heavy top quark limit, while
the solid NNLO line includes the exact scale dependence at NNLO, which has
been obtained from the exact NLO result by means of renormalization group
methods \cite{hresum}. The full NNLO curve has been normalized to the
$\gamma_2$ curve at
$\mu = M = M_H$. Fig.~\ref{fg:Higgs_sc} clearly indicates a strong stabilization
of the scale dependence at NNLO, which develops a broad maximum around the
natural scale $\mu=M\sim M_H$ in contrast with the NLO curve, which is a
monotonic function of the scales. Thus with NNLO parton densities it would be
possible to predict the Higgs production cross section at the LHC accurately
with small theoretical uncertainties from the remaining scale dependence.
However, the non-perturbative uncertainties related to the renormalon
singularity of the full resummed result may be sizeable.

\section{Acknowledgements}

I would like to thank the organizers, especially Bernd Kniehl, for the
encouraging atmosphere at the workshop. Moreover, it is a pleasure to thank
my collaborators Abdelhak Djouadi, Sally Dawson, Eric Laenen and Michael
Kr\"amer in the new work presented here.


\section{References}

\begin{thebibliography}{99}

\bibitem{S:higgs} P.W.~Higgs, Phys.~Rev.~Lett.~{\bf 12} (1964) 132 and
Phys.~Rev.~{\bf 145} (1966) 1156; 
F.~Englert and R.~Brout, Phys.~Rev.~Lett.~{\bf 13} (1964) 321;
G.S.~Guralnik,  C.R.~Hagen and  T.W.~Kibble, Phys.~Rev.~Lett.~{\bf 13} (1964)
585.

\bibitem{S:lep1} J.-F.~Grivaz, Proceedings, International Europhysics
Conference on High Energy Physics, Brussels, 1995.

\bibitem{S:lattice} N.~Cabibbo, L.~Maiani, G.~Parisi and R.~Petronzio,
Nucl.~Phys.~{\bf B158} (1979) 295;
M.~Chanowitz, M.~Furman and I.~Hinchliffe, Phys.~Lett.~{\bf B78} (1978) 285;
R.A.~Flores and M.~Sher, Phys.~Rev.~{\bf D27} (1983) 1679;
M.~Lindner, Z.~Phys.~{\bf C31} (1986) 295;
M.~Sher, Phys.~Rep.~{\bf 179} (1989) 273;
A.~Hasenfratz, K.~Jansen, C.~Lang, T.~Neuhaus and H.~Yoneyama,
Phys.~Lett.~{\bf B199} (1987) 531;
J.~Kuti, L.~Liu and Y. Shen, Phys.~Rev.~Lett.~{\bf 61} (1988) 678;
M.~L\"uscher and P.~Weisz, Nucl.~Phys.~{\bf B318} (1989) 705. 

\bibitem{S:glufus} H.~Georgi, S.~Glashow, M.~Machacek and D.V.~Nanopoulos,
Phys.\ Rev.\ Lett.\ {\bf 40} (1978) 692.

\bibitem{SDGZ}\label{SDGZ}
D.~Graudenz, M.~Spira and P.~M.~Zerwas, Phys.~Rev.~Lett.~{\bf 70} (1993)
1372;
M.~Spira, A.~Djouadi, D.~Graudenz and P.~M.~Zerwas, Phys.~Lett.~{\bf B318}
(1993) 347 and \np{453}{95}{17}.

\bibitem{S:limit} A.~Djouadi, M.~Spira and P.M.~Zerwas, Phys.~Lett.~{\bf
B264} (1991) 440;
S.\ Dawson, Nucl.\ Phys.\ {\bf B359} (1991) 283;
R.P.\ Kauffman and W.\ Schaffer, Phys.\ Rev.\ {\bf D49} (1994) 551;
S.\ Dawson and R.P.\ Kauffman, Phys.\ Rev.\ {\bf D49} (1994) 2298. 

\bibitem{S:hbound}
Y.~Okada, M.~Yamaguchi and T.~Yanagida, Prog.~Theor.~Phys.~{\bf 85} (1991) 1;
H.~Haber and R.~Hempfling, Phys.~Rev.~Lett.~{\bf 66} (1991) 1815;
J.~Ellis, G.~Ridolfi and F.~Zwirner, Phys.~Lett.~{\bf B257} (1991) 83;
R.~Barbieri, F.~Caravaglios and M.~Frigeni, Phys.~Lett.~{\bf B258} (1991) 167;
M.~Carena, J.~Espinosa, M.~Quiros and C.E.M.~Wagner, Phys.~Lett.~{\bf B355}
(1995) 209;
M.~Carena, M.~Quiros and C.E.M.~Wagner, Nucl.~Phys.~{\bf B461} (1996) 407.

\bibitem{S:phibb} Z.~Kunszt and F.~Zwirner, Nucl.~Phys.~{\bf B385} (1992) 3;
V.~Barger, M.~Berger, S.~Stange and R.~Phillips, Phys.~Rev.~{\bf D45} (1992)
4128;
H.~Baer, M.~Bisset, C.~Kao and X.~Tata, Phys.~Rev.~{\bf D46} (1992) 1067;
J.~F.~Gunion and L.~Orr, Phys.~Rev.~{\bf D46} (1992) 2052;
J.~F.~Gunion, H.~E.~Haber and C.~Kao, Phys.~Rev.~{\bf D46} (1992) 2907.

\bibitem{S:squark} S.~Dawson, A.~Djouadi and M.~Spira,
Phys.~Rev.~Lett.~{\bf 77} (1996) 16.

\bibitem{lowen}
J.\ Ellis, M.K.\ Gaillard and D.V.\ Nanopoulos, Nucl.\ Phys.\ {\bf B106}
(1976) 292;
A.~Vainshtein, M.~Voloshin, V.~Zakharov and
M.~Shifman, Sov.\ J.\ Nucl.\ Phys.\ {\bf 30} (1979) 711; 
B.~Kniehl and M.~Spira, Z.~Phys.\ {\bf C69} (1995) 77.
 
\bibitem{susybet}
L.~Alvarez-Gaum\'e, J.~Polchinski, and M.~Wise, Nucl.\ Phys.\ {\bf B221}
(1983) 495;
S.~Martin and M.~Vaughn, Phys.\ Rev.\ {\bf D50} (1994) 2282;
J.~Derendinger and C. Savoy, Nucl.~Phys.\ {\bf B237} (1984) 307.

\bibitem{betafun}
W.~Caswell, Phys.~Rev.~Lett.~{\bf 33} (1974) 244;
D.R.T. Jones, Phys.\ Rev.\ {\bf D25} (1982) 581;
M. Einhorn and D.R.T. Jones, Nucl.\ Phys.\ {\bf B196} (1982) 475. 

\bibitem{GRV} M.~Gl\"uck, E.~Reya and A.~Vogt, Z.~Phys.~{\bf C53} (1992) 127.

\bibitem{CTEQ4}\label{CTEQ}
H.L. Lai et al. (CTEQ Collab.), Phys.\ Rev.\ {\bf D55} (1997) 1280.

\bibitem{CLS}\label{CLS}
H.~Contopanagos, E.~Laenen and G.~Sterman, \np{484}{97}{303}.

\bibitem{hresum}
E.\ Laenen, M.\ Kr\"amer and M.\ Spira, Report CERN-TH/96-231, hep-ph/9611272.

\bibitem{bernwetz} W.~Bernreuther and W.~Wetzel, Nucl.\ Phys.\ {\bf B197} (1982)
228;
W.~Bernreuther, Ann.~Phys.~{\bf 151} (1983) 127;
S.A.\ Larin, T.\ van Ritbergen and J.A.M.\ Vermaseren, Nucl.\ Phys.\
{\bf B438} (1995) 278.

\bibitem{factheo}
G.\ Sterman, Nucl.\ Phys.\ {\bf B281} (1987) 310.

\bibitem{renorm}
D.\ Appell, P.\ Mackenzie and G.\ Sterman, Nucl.\ Phys.\ {\bf B309} (1988) 259;
H.\ Contopanagos and G.\ Sterman, Nucl.\ Phys.\ {\bf B419} (1994) 77;
Y.\ Dokshitser, G.\ Marchesini and B.\ Webber, Report CERN-TH/95-281,
hep-ph/9512336;
S.\ Catani, M.\ Mangano, P.\ Nason and L.\ Trentadue, Nucl.\ Phys.\ {\bf B478}
(1996) 273. 

\end{thebibliography}
\end{document}